\documentclass[prd,twocolumn,nofootinbib,preprintnumbers]{revtex4}
\usepackage[dvips]{graphicx}
\usepackage{bm,latexsym,amsmath,amssymb,amsfonts}

\newcommand\beq{\begin{eqnarray}}

\begin{document}

\title{Cosmological Kaluza-Klein branes in black brane spacetimes}

\author{Masato~Minamitsuji}
\email[Email: ]{masato.minamitsuji"at"physik.uni-muenchen.de}
\affiliation{Arnold-Sommerfeld-Center for Theoretical Physics, Department f\"{u}r Physik, Ludwig-Maximilians-Universit\"{a}t, Theresienstr. 37, D-80333, Munich, Germany}

\begin{abstract} 
We discuss the comsological evolution of a brane in the $D(>6)$-dimensional
black brane spacetime in the context of the Kaluza-Klein (KK)
braneworld scheme, i.e., to consider KK compactification on the brane.
The bulk spacetime is composed of two copies of a patch of
$D$-dimensional black three-brane solution.
The near-horizon geometry is given by
$AdS_{5}\times S^{(D-5)}$ while in the asymptotic infinity the spacetime
approaches $D$-dimensional Minkowski.
We consider the brane motion from the near-horizon region toward the spatial
infinity, which induces cosmology on the brane.
As is expected,
in the early times, namely when the brane is located in the near-horizon region,
the effective cosmology on the brane coincides with that in the second
Randall-Sundrum (RS II) model.
Then, the brane cosmology starts to deviate from the RS type one
since the dynamics of KK compactified dimensions becomes significant.
We find that the brane Universe cannot reach the asymptotic infinity,
irrespectively of the components of matter on the brane.
\end{abstract}

\preprint{LMU-ASC 29/08}
\maketitle

Since the stimulating proposals by
Randall and Sundrum (RS) \cite{rs2,rs1},
the cosmological aspects of braneworld models
have been studied in various literature \cite{rev}.
In the second RS (RS II) model \cite{rs2},
at the low energy scales
the four-dimensional cosmology is recovered on the brane
due to the strong warping of the extra dimension.
Then, six and higher-dimensional models
have started to attract growing interests.
In higher dimensions, there will be various types of braneworld models,
for example, those with flux-stabilized compact extra dimensions,
with brane intersections.
Codimension-two brane models with flux-stabilized extra dimensions
have been originally focused on as a simple realization of 
large extra dimensions \cite{add} 
and as a way of resolution of the cosmological constant problem \cite{cc+}
(see, however, e.g., \cite{cc-}).
Recently, these six dimensional models have been discussed
from the different aspects. 
It is well-known that
a four-dimensional defect in six or higher dimensions
generically has a problem on localization of ordinary matter
on the brane due to the stronger self-gravity. 
Ways to circumvent this problem have to be developed.
In six dimensional models, the ways of brane regularization to study
gravity and cosmology on the brane
have been studied in \cite{pst,vc,thick6d}.
The way employed in Ref. \cite{pst,thick6d}
is that the original codimension-two brane is replaced
with a ring-like codimension-one brane wrapped around the axis of 
symmetry of the bulk.

Our main purpose is to construct brane cosmological models
in higher dimensional spacetime.
In order to do so,
one would consider regularizations of branes with
higher-codimensions as
the extensions of ways developed in the studies of
six-dimensional (codimension-two) brane models.
But, in fact, it seems to be difficult to construct
the braneworld models in such a way,
since these branes have stronger self-gravity and develop severe singularities.
Thus, in this article we take a (similar but) different approach,
instead of regularizing a brane with higher codimensions.
We focus on the hybrid construction of the Kaluza-Klein (KK)
and braneworld compactifications, i.e.,
to consider KK compactifications on the brane.
Such a way of construction is called {\it Kaluza-Klein braneworld} \cite{kkbw}.
Actually, the way of regularizations deveoloped in \cite{pst,thick6d}
can be seen as a kind of KK braneworlds.
In the context of six dimensions,
such a hybrid construction has been presented in Ref. \cite{hb},
before the works \cite{pst,thick6d}.
In their construction, two identical copies of a regular
and compact internal space are glued at the position of
the brane and clearly the resultant codimension-one brane
cannot be interpreted as a {\it regularized} object.
The scheme of KK braneworlds could be one of the most powerful
tools to construct successful braneworld models in higher dimensions.
\footnote{See for several attempts
to realize the localization of ordinary matter and gravity on the 
brane in higher dimensions, e.g.,
by constructing regular solitonic solutions \cite{hcodim},
by replacing the brane with the core region where matter is distributed smoothly \cite{regular},
by adding higher curvature terms to the theory \cite{hcodim2}
and by considering brane intersections \cite{insc}.}
General properties of KK braneworlds have been investigated recently,
e.g., in Ref. \cite{kkbw}. 
Cosmology on a KK brane can be realized by considering the motion of
the KK brane into the bulk spacetime.
In this article, we construct an explicit model of cosmological
Kaluza-Klein braneworld in the higher-dimensional spacetime.

In constructing KK cosmological brane models,
the important problem is
how one makes the KK compactified dimensions
to be invisible to the obserbers on the brane.
In this article, we consider the black brane solutions,
whose near horzion geometry is $AdS_{5}\times S^{(D-5)}$
while in the asymptotic infinity the spacetime
approaches $D$-dimensional Minkowski.
We consider the motion of the brane from the near horizon 
toward the asymptotic infinity.
Thus, in the early times the brane stays in the near-horizon region.
In the black brane spacetime, 
the magnetic $(D-5)$-form field keeps the 
the size of sphercal compact dimensions enough compact
in the near-horizon region
and helps to make the dynamics of these dimensions
to be invisible on the brane in the early times.
In addition,
the spacetime has the higy warped extra dimension.
The well-behaved cosmological feature of the RS II model
is due to the realization of the warped structure of the
extra dimension.
Such a warped structure of extra dimensions also would help
for the standard four-dimensional cosmology to be recovered on the brane
due to the localization of the zero mode gravitions.

Finally, it is expected that in the later times,
the brane cosmology starts to deviate from
the RS II type one because the dynamics of the KK
compactified dimensions could be significaint.
We discuss this point in terms of the effective potential
for the scale factor.

{\it Cosmological Kaluza-Klein branes:}

We briefly review the cosmology on a $(D-1)$-dimensional
KK brane in $D$-dimensional spacetime.
We follow the well-established method developed in Ref. \cite{5ddw}.
We consider a motion of a brane in the static background.
We start from the following static $D(>6)$-dimensional metric ansatz:
\begin{eqnarray}
 ds^2&=&-F(r)^2 dt^2+A(r)^2 \delta_{ij}dx^i dx^j
\nonumber\\
&+&B(r)^2 dr^2+ C(r)^2 h_{ab}dy^a dy^b\,. \label{gold}
\end{eqnarray}
We assume the $Z_2$ symmetry with respect to the brane.
$h_{ab}$ represents the metric of $(D-5)$-dimensional compact space.
The indices $i(=1,2,3)$ and $a$ run over the ordinary three-dimensional
space and $(D-5)$-dimensional compact space, respectly.
The brane proper time $\tau$ and induced metric are given by
\begin{eqnarray}
&& -F\big(r(\tau)\big)^2\dot{t}^2
+B\big(r(\tau)\big)^2\dot{r}^2=-1\,,
\nonumber\\
&&ds^2_{\rm ind}
=-d\tau^2+a^2 \delta_{ij}dx^i dx^j
+C(a)^2 h_{ab}dy^a dy^b\,.
\end{eqnarray}
where the cosmological scale factor is given by
$a=A\big|_{r=r_b}$ ($r_b$ is the radial position
of the brane) and {\it dot} means the derivative
with respect to the brane proper time $\tau$.

Then, we derive the equations of motion of the brane in the static bulk.
The non-trivial
components of Israel junction condition are given by
\begin{eqnarray}
&& \Big(\frac{3}{a}+\frac{(D-5) C_{,a}}{C}\Big)
\sqrt{\tilde B^{-2}+\dot{a}^2}
=\frac{\rho}{2M_D^{D-2}},\label{rho}
\\
&&\frac{\ddot{a}+\frac{\tilde B_{,a}}{\tilde B}\dot{a}^2}
      {\sqrt{\tilde B^{-2}+\dot{a}^2}}
+\Big(
  \frac{F_{,a}}{F}
  -\frac{1}{a}
 \Big)\sqrt{\tilde B^{-2}+\dot{a}^2}
\nonumber\\
&=&-\frac{1}{2 M^{D-2}_{D}}\Big(\rho+p\Big),
\label{acc}
\\
&&\frac{\ddot{a}+\frac{\tilde B_{,a}}{\tilde B}\dot{a}^2}
      {\sqrt{\tilde B^{-2}+\dot{a}^2}}
+\Big(
  \frac{F_{,a}}{F}
  -\frac{C_{,a}}{C}
 \Big)\sqrt{\tilde B^{-2}+\dot{a}^2}
\nonumber\\
&=&-\frac{1}{2 M^{D-2}_D}\Big(\rho+q\Big)\label{acc2},
\end{eqnarray}
where $\tilde B da= B dr$
and $M_{D}$ represents the $D$-dimensional gravitational scale.
$\rho$, $q$ and $p$ are ($(D-5)$-dimensional) energy density, pressure along
the internal dimensions
and the one along the external dimensions, respectively.
If there are dynamical degrees of freedom than the metric in the bulk,
we need to take their jump conditions into account.
The first component of the Israel junction conditions Eq. (\ref{rho})
gives the effective Friedmann equation
\begin{eqnarray}
 \Big(\frac{\dot{a}}{a}\Big)^2=
\frac{\rho^2}
      {4M_{D}^{2(D-2)}\big(3+\frac{(D-5) a}{C}\frac{dC}{da}\big)^2}
-\frac{1}{a^2 {\tilde B}^2}. \label{rf}
\end{eqnarray}

Before closing this section, we mention
the other well-known approach, i.e., to consider a fixed brane 
in the time-dependent bulk spacetime which was employed e.g., in \cite{bdl}.
In this approach, the effective cosmological equations are obtained
by integrating the bulk gravitational equations.
Of course, these two pictures must be equivalent.
In the RS II model, in this picture
an integration constant appears in the effective cosmological equations.
This integration constant corresponds to the
mass of the static bulk black hole in our picture
(the parameter $\mu$ in our case which appears in the next section).


{\it Black three-brane solutions:}

As one of the simplest higher-dimensional solutions,
we consider the $D$-dimensional black three-brane solutions (see e.g., \cite{bb}).
We start from the following $D$-dimensional theory:
\begin{eqnarray}
 S=\frac{M_{D}^{D-2}}{2}
\int d^{D}X\sqrt{-G}
\Big(
R[G]-\frac{1}{2(D-5)!}F_{[D-5]}^2
\Big),
\end{eqnarray}
where $G_{AB}$ and $R[G]$ are the $D$-dimensional metric and the
Ricci scalar associated with the metric $G_{AB}$, respectively.
$F_{[D-5]}$ represents the $(D-5)$-form field strength.
This theory contains a series of the black three-brane solutions,
whose metric is given by
\begin{eqnarray}
 ds^2&=&H^{-1/2}\Big(-f dt^2+\delta_{ij}dx^i dx^j\Big)
\nonumber\\
&+&H^{2/(D-6)}
\Big(f^{-1}dr^2+r^2 h_{ab}dy^ady^b\Big),\label{bb}
\end{eqnarray}
where $\delta_{ij}$ and $h_{ab}$ are metric of 
flat three-dimensional space and $(D-5)$-sphere,
respectively.
We also define
\begin{eqnarray}
 H(r):=1+\frac{Q}{r^{D-6}},\quad
 f(r)=1-2\frac{\mu}{r^{D-6}},
\end{eqnarray}
and $h_{ab}$ is the metric of $(D-5)$-sphere.
$\mu$ and $Q$ are parameters of the solutions.
We assume that the function $f$ is positive.
Thus, the radial coordinate $r$ has the minimal value at
the horizon position $r=r_{\rm min}:=(2\mu)^{1/(D-6)}$.
The case that $\mu=0$
corresponds to the extremal solutions.
The near horizon geometry is ${\rm AdS}_{5}\times S^{(D-5)}$.
The effective curvature radius of AdS spacetime is
related to
the parameter $Q$
\begin{eqnarray}
\ell:=\frac{4}{D-6}Q^{1/(D-6)}.\label{ell}
\end{eqnarray}
The $(D-5)$-form field
acts on $S^{(D-5)}$.
The magnetic $(D-5)$-form field strength is given by
\begin{eqnarray}
F^{y^1\cdots y^{(D-5)}}
=\frac{1}{\sqrt{-G}}
 \epsilon^{y^1\cdots y^{D-5}}
 E'(r)
\end{eqnarray}
where
\begin{eqnarray}
E(r)=\sqrt{\frac{Q}{Q+2\mu}\frac{D-2}{2(D-6)}}
      \frac{f(r)}{H(r)}\,.
\end{eqnarray}
At the spatial infinity $r\to \infty$, the spacetime approaches
$D$-dimensional Minkowski.
In the case $D=10$, the solution corresponds
to a stack of coincident D3-branes at the low energy scales.


The reason to consider the black brane spacetime is as follows.
Firstly, the spacetime contains the warped spatial dimension.
The well-behaved cosmological feature of the RS II model
is due to the realization of the warped structure of the
extra dimension.
Such a warped structure of extra dimensions may also help
to construct the realistic brane cosmological models in higher dimensions.
Also, in string compactifications,
the highly warped structure is induced
gravitationally by the branes localized at a certain place in the internal space.
The black brane spacetime is one of the simplest examples 
of such spacetimes.
Secondly,
in in the near-horizon of the black brane spacetimes, 
the magnetic $(D-5)$-form field keeps spherical dimensions 
compact enough.

We apply the previously mentioned KK braneworld scheme
to the black brane spacetime.
The bulk geometry is constructed by the standard cut-and-paste procedure.
We excise the region $r>r_b$ from the original black brane spacetime, 
where $r_b$ is a certain radial position, and
glue two copies of the remaining spacetime at $r=r_b$.
Then, two copies of the patch $r_{\rm min}<r<r_b$ are bounded by a codimension-one
brane at $r=r_b$.
The compact spatial dimensions along $(D-5)$-sphere
(as well as the ordinary four-dimensional spacetime dimensions)
are involved into the brane worldvolume as KK compactified dimensions.
We identify two copies of the bulk by 
imposing the reflection ($Z_2$-)symmetry with respect to the KK brane.
An expanding cosmology is realized as a brane motion from the spatial black brane
horizon toward the spatial infinity.


{\it Cosmology on the KK brane:}

The bulk metric functions $A$, $B$, $C$
and $F$ in Eq. (\ref{gold}) are now
\begin{eqnarray}
&& A=\frac{1}{H^{1/4}},\quad B=\Big(\frac{f^{1/2}}{H^{1/(D-6)}}\Big)^{-1},
\nonumber\\
&& C= r H^{1/(D-6)},\quad F=\frac{f^{1/2}}{H^{1/4}}.
\end{eqnarray}
The cosmic scale factor is given by
$a(\tau):=A(r_b)$.
For the nonzero $\mu$, 
the cosmic scale factor has non-zero minimal value:
\begin{eqnarray}
a_{\rm min}
=\Big(\frac{2\mu/Q}{1+2\mu/Q}\Big)^{1/4},
\end{eqnarray}
which corresponds to $r=r_{\rm min}$.
On the other hand, in any case,
the cosmic scale factor also has the maximal size $a=1$ 
for $r\to \infty$.
However, as we will see the brane Universe cannot reach $a=1$.
In order to have enough cosmological expansions,
we assume the near-extremal condition $\mu/Q\ll 1$.
The physical size of KK directions on the brane is given by
$C=Q^{1/(D-6)} /(1-a^4)^{1/(D-6)}$ and thus in the near horizon region $a\ll1$
is as small as the effective AdS radius $\ell$.

Then, by integrating over the $S^{D-5}$ dimensions on the brane,
the effective four-dimensional energy density is
obtained as
\begin{eqnarray}
 \Sigma
&:=&\big(C(a)\big)^{D-5}
\int d^{D-5}y\,\sqrt{{\rm det}(h_{ab})}\, \rho
\nonumber\\
&=&\rho\frac{Q^{(D-5)/(D-6)}
                  \Omega_{(D-5)}}
            {(1-a^4)^{(D-5)/(D-6)}}\,,
\label{eff_e}
\end{eqnarray}
where $h_{ab}$
represents the metric of the $(D-5)$-sphere
with the unit radius and
$\Omega_{D-5}=\int d^{D-5}y\,\sqrt{{\rm det}(h_{ab})}
=\pi^{\frac{D-5}{2}}/\Gamma\big(1+\frac{D-5}{2}\big)$
is its volume.
From the Eq. (\ref{rf}),
the effective Friedmann equation becomes
\begin{widetext}
\begin{eqnarray}
\Big(\frac{\dot{a}}{a}\Big)^2
&=& \frac{1}{4Q^{\frac{2(D-5)}{(D-6)}} \Omega_{D-5}^2M_D^{2(D-2)}}
\frac{(1-a^4)^{2+\frac{2(D-5)}{D-6}}(D-6)^2\Sigma^2}
     {\big[3(D-6)(1-a^4)+4(D-5)a^4\big]^2}
-\Big(\frac{D-6}{4}\Big)^2
\frac{(1-a^4)^{2+\frac{2}{D-6}}
      \big[(1+\frac{2\mu}{Q})a^4-\frac{2\mu}{Q}\big] }
     {a^4 Q^{2/(D-6)}}\label{akkin}
\end{eqnarray}
\end{widetext}
In the early times $a\ll 1$, the Eq. (\ref{akkin}) is reduced to
the form
\begin{eqnarray}
\Big(\frac{\dot{a}}{a}\Big)^2
&\approx &
\frac{1}{36 Q^{\frac{2(D-5)}{(D-6)}}\Omega_{D-5}^2M_D^{2(D-2)}}
\Sigma^2
\nonumber\\
&-&\Big(\frac{D-6}{4}\Big)^2\frac{1+\frac{2\mu}{Q}}{Q^{2/(D-6)}}
\nonumber \\
&-&\Big(\frac{D-6}{4}\Big)^2\frac{2\mu}{a^4Q^{1+2/(D-6)}}.
\label{early}
\end{eqnarray}
Here, it is useful to decompose
the effective energy density $\Sigma$ into the constant and time-dependent parts as
$\Sigma=\Sigma_0+\Sigma_1$,
where $\Sigma_0$ is the constant part of the effective
brane energy density and $\Sigma_1$ is the remaining time-dependent part.
The constant part of the brane energy density
can also be decomposed into two parts $\Sigma_0=\Sigma_{\rm RS}+\delta$, 
where $\Sigma_0$ is chosen as
\begin{eqnarray}
 \Sigma_{\rm RS}:=6\Big(\frac{D-6}{4}\Big) M_{D}^{D-2} Q \Omega_{(D-5)}
         \Big(1+\frac{2\mu}{Q}\Big)^{1/2}. \label{RS}
\end{eqnarray}
In the case that $\delta=0$ , it is clear that the constant part in the right-hand-side of 
the Eq. (\ref{early}) vanishes.
In the early times,
the effective Friedmann equation given by Eq. (\ref{early})
is rewritten as
\begin{eqnarray}
 \Big(\frac{\dot{a}}{a}\Big)^2
&\approx&
\frac{1}{3}\lambda_{\rm eff}
+\frac{1}{3M_4^2}
\Big(\Sigma_1+\Sigma_{\rm DR}\Big)
+\frac{1+\frac{2\mu}{Q}}{\ell^2}
\Big(\frac{\Sigma_1}{\Sigma_0}\Big)^2,
\end{eqnarray}
where the effective gravitational scale $M_4$ and
the effective cosmological constant on the brane $\lambda_{\rm eff}$
are given by
\begin{eqnarray}
 M_4^2
:=\frac{4}{D-6}
  \frac{M^{D-2}_D Q^{(D-4)/(D-6)}\Omega_{D-5}}
       {\big(1+\frac{2\mu}{Q}\big)^{1/2}
        \big(1+\frac{\delta}{\Sigma_{\rm RS}}\big)}.
\label{pl2}
\end{eqnarray}
and
\begin{eqnarray}
\frac{\lambda_{\rm eff}}{3}
:=
\ell^{-2}
\Big(1+\frac{2\mu}{Q}\Big)
\Big(
 \frac{2\delta}{\Sigma_{\rm RS}}
+\frac{\delta^2}{\Sigma_{\rm RS}^2}
\Big),
\end{eqnarray}
respectively.
We also define the effective energy density:
\begin{eqnarray}
\Sigma_{\rm DR}
:=3M_4^2 \Big(\frac{D-6}{4}\Big)^2
\frac{2\mu}{Q^{1+2/(D-6)}a^4}.\label{dr2}
\end{eqnarray}
The result is nothing but the one
in the RS II model \cite{rev}.
Especially,
in the low energy regime $\Sigma_{1}\ll\Sigma_{\rm RS}$,
the standard cosmology is recovered except for the term
$\Sigma_{\rm DR}$.
By combining Eq. (\ref{RS}) with (\ref{pl2}), we find
\begin{eqnarray}
\Sigma_{\rm RS}&\simeq&
\frac{6M_4^2}{\ell^2}
\Big(1+\frac{2\mu}{Q}\Big).
 \label{RS2}
\end{eqnarray}
As is seen previously,
in the case that $\delta=0$, as in the RS  II model \cite{rs2},
the effective cosmological constant $\lambda_{\rm eff}$ vanishes.
The deviation from the extremal condition $\mu\neq 0$
gives rise to the so-called {\it dark radiation}
(see Ref.\cite{rev}),
given by $\Sigma_{\rm DR}$.
In the RS II cosmology, the dark radiation arises
due to the presence of a black hole in the bulk.
In the current case, similarly the near-horizon geometry is approximately given by the product of 
the five-dimensional Schwarzdschild-AdS with $S^{(D-5)}$.
As we will see later, the pressure equation in the corresponding bulk region 
also has the same structure as in the RS II cosmology.

In the later times, cosmology could deviate from the RS one
because the dynamics of the KK compactified dimensions
becomes important.
To see this, it is useful to rewrite the effective Friedmann
equation into the form in an analogy with the classical mechanics as
$\dot{a}^2+U(a)=0$.
For instance in the extremal case $\mu=0$,
the potential term is given by
\begin{eqnarray}
U(a)
&:=&-
\ell^{-2}a^2 (1-a^4)^{2+2/(D-6)}
\nonumber\\
&\times&\Big[
\frac{(1-a^4)^2\big(1+\frac{\delta}{\Sigma_{\rm RS}}+\frac{\Sigma_1}{\Sigma_{\rm RS}}\big)^2}
     {\big(1+\frac{(D-2)}{3(D-6)}a^4\big)^2}
-1
\Big].
\end{eqnarray}
As easily seen, the potential term becomes positive before arriving
the maximal value $a=a_{\rm max}$.
In the case of the vacuum brane $\Sigma_1=0$, this maximal value of
cosmic scale factor is given by
\begin{eqnarray}
a_{\rm max}
=\left(\frac{\delta}{\delta+\frac{4(D-5)}{3(D-6)}\Sigma_{\rm RS}}
\right)^{1/4}
\end{eqnarray}
In Fig. 1 and 2, the typical behaviors of the potential $U(a)$
without the ordinary matter $\Sigma_1=0$ and 
with dust fluid on the brane are shown, respectively.
The function $U(a)$ is vanishes at $a=a_{\rm max}$.
After the brane reaches the maximal size, the brane goes back toward $a\to 0$.
Similar behaviors can be seen in the near-extremal case. 
Thus, we find that {\it the brane Universe cannot reach the asymptotic infinity,
irrespectively of the components of the matter on the brane}.
In other words,
observers on the brane never see the KK compactified extra dimensions.
The result implies that
the cosmological brane cannot escape from the gravitational potential
induced by the black brane horizon.
\begin{figure} 
\begin{minipage}[t]{.40\textwidth}
   \begin{center}
    \includegraphics[scale=.80]{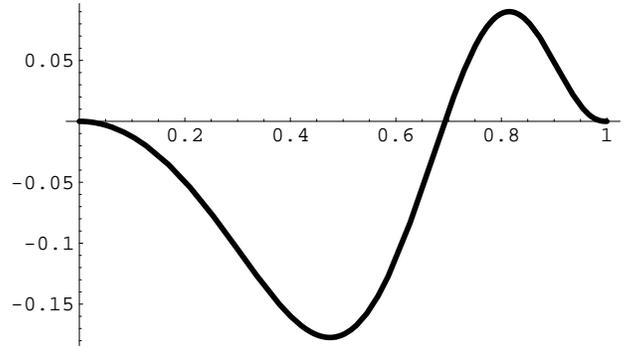}
        \caption{ The function $U(a)$
is shown as a function of $a$ for $D=10$, $\delta/\Sigma_{\rm RS}=0.5$ and $\Sigma_1=0$.}
   \end{center}
   \end{minipage}
\end{figure}
\begin{figure} 
\begin{minipage}[t]{.40\textwidth}
   \begin{center}
    \includegraphics[scale=.80]{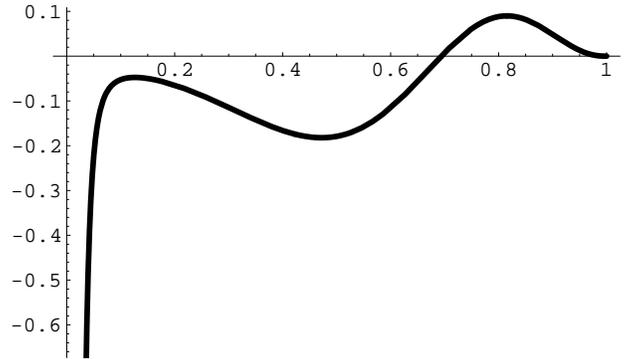}
        \caption{ The function $U(a)$
is shown as a function of $a$ for $D=10$, $\delta/\Sigma_{\rm RS}=0.5$. 
$\Sigma_1/\Sigma_0=0.005/a^3$ is chosen as dust fluid.}
   \end{center}
   \end{minipage}
\end{figure}


As previously mentioned, in the special case that $D=10$,
the black three-brane solution corresponds to
a low energy description of a stack of $N$ coincident (BPS for $\mu=0$) D3-branes in type IIB string theory.
These parameters in the solution can be expressed as
$M_{10}^8=2/(2\pi)^7 g_s^2 {\alpha{}'}^4$
and
$Q=\ell^4=4 \pi N g_s {\alpha{}'}^2$
where $\ell_s=(\alpha{}')^{1/2}$ and $g_s$
represent string length scale and string coupling constant,
respectively.
Note that
the supergravity description of D3-branes
is valid in the case that the effective AdS curvature scale $\ell=Q^{1/4}$ is 
much larger than $\ell_s$ (hence $Ng_s\gg 1 $) and
quantum gravitational corrections are negligible,
 i.e., $\ell\gg M_{10}^{-1}$ (hence $N\gg 1$).
Thus, assuming the near-extremal condition $\mu/Q\ll1$ and
an optimistic choice
of the brane tension $\delta/\Sigma_{\rm RS}=O(1)$,
the effective four-dimensional quantities are also written as 
\begin{eqnarray}
M_4^2\simeq \frac{N^{3/2}}
          {8\pi^{5/2}g_s^{1/2} {\alpha'}},
\quad
\Sigma_{\rm RS}
\simeq \frac{3N}{8\pi^3 g_s {\alpha'}^2}. \label{string}
\end{eqnarray}
Although our approach is higher-dimensional,
in the sense that we start from the original higher-dimensional theory
and then secondly integrate over the internal dimensions after obtaining
the junction conditions,
we now confirm that
in the early time the brane cosmology obtained in our approach 
is equivalent to the one in the approach in the effective theory approach
in which the internal dimensions are integrated over firstly.

{\it Experimental and observational bounds:}

As in the cases of the RS II cosmology (see e.g.,\cite{rs_cosmo}), 
several experimental and observational bounds are obtained.
A lower bound on the RS brane tension is given by the condition that
the density square term in the effective Friedmann equation must 
be negligible
before the Big Bang nucleosynthesis (BBN) $\Sigma_{\rm RS}> (1 {\rm MeV})^4$.
But a more stringent bound is obtained from the results of the table top tests
on Newton's law, implying $\ell < 0.1 {\rm mm}$ and hence $\Sigma_{\rm RS} > \big(10^3 {\rm GeV}\big)^4$ 
in the assumption that $\mu/Q\ll 1$ since $M_4\approx 10^{18}{\rm GeV}$.

In the ten-dimensional ($D=10$) case,
from the Eq. (\ref{string}),
we find
\begin{eqnarray}
N^2\simeq \frac{24\pi^2 M_4^4}{\Sigma_{\rm RS}},\quad
g_s{\alpha'}^2
\simeq
\Big(\frac{3}{2}\Big)^{3/2}
\frac{M_4^2}{\pi^2\Sigma_{\rm RS}^{3/2}}.
\end{eqnarray}
The condition that $\Sigma_{\rm RS} > (10^3 {\rm GeV})^4$
(hence $\ell<0.1{\rm mm}$) implies $N< 10^{30}$ and
$M_{10}\sim \big(g_s {\alpha'}^2\big)^{-1/8} >  10^{-5} {\rm GeV}$.
Thus,
there is a large parameter space which is consistent with the 
conditions that $N g_s\gg 1$ and $N\gg 1$ for which the classical description
of the solution
is valid.
Note that $\big(\ell/0.1{\rm mm}\big)\simeq \big(N/10^{30}\big)$,
since $N\simeq 2\pi M_4\ell$.

As we have seen,
in the near-horizon region
the effective cosmology on the brane is well approximated by
the one in the RS II model.
According to the {\it AdS/conformal field theory (CFT)} (
more precisely $AdS_{5}/ CFT_4$ ) {\it correspondence} \cite{adscft},
which states that
in type IIB string theory
the (super)gravity theory in the five-dimensional AdS spacetime
is equivalent to the ${\cal N}=4$ super-conformal $U(N)$ Yang-Mills theory at the boundary,
gravity in RS II braneworld seems to be equivalent to the four-dimensional
gravity coupled to CFT on the brane.
There are several examples in which the equivalence (up to $O(\ell^{2})$)
has been confirmed \cite{adscft_bw}, including
the case of cosmological branes.
Such an equivalence should be valid for our cosmological brane. 
The five-dimensional effective gravitational scale is related to the four-dimensional one
$M_{5}^3\approx M_4^2/\ell$.
Then, the CFT parameters are related to the five-dimensional gravitational
ones.
For instance, the CFT degrees of freedom is given by $N^2 \approx (2\pi)^2 M_5^3\ell^3$.
Since $\ell^{-1}$ and $N^2$ are the cut-off energy scale
and the degrees of freedom of CFT, respectively.
The relation $M_4^2\simeq N^2 \ell^{-2}$ can be seen as a typical example
that the bound $M_4^2\gtrsim n \Lambda^2$ (discussed in Ref. \cite{dvali})
is saturated. 
Here, $n$ is the number of species of quantum fields with mass scale $\Lambda$.

Inspired by the AdS/CFT correspondence, it has been conjectured
that a black hole on a RS II brane is classically unstable \cite{cbh}.
Such a conjecture is based on the argument
that the a five dimensional black hole localized on the brane should be
equivalent to the four-dimensional quantum-corrected black hole
of the gravity theory coupled to CFT.
From the CFT point of view,
such a brane black hole could decay into large ($\sim N^2$) CFT degrees of freedom
and its lifetime $\tau_{\rm BH}$ should be much shorter than that of standard one:
$\tau_{\rm BH}\simeq 10^4 (M_{\rm BH}/M_{\odot})^3 (0.1 {\rm mm}/\ell)^2 {\rm year}$
(or equivalently,
$\tau_{\rm BH}\simeq 10^4 (M_{\rm BH}/M_{\odot})^3 (10^{30}/N)^2 {\rm year}$
) \cite{cbh},
where $M_{\rm BH}$ and $M_{\odot}$ are the masses of a black hole and the Sun,
respectively.
If the conjecture in RS II model is true, of course this must be true in our
construction (as long as the brane stays in the near horizon region).
This could give a stronger bound on $\ell$.
For example, in order for a black hole with solar mass $M_{\rm BH}\simeq M_{\odot}$,
to survive today, $\tau_{\rm BH}$ must be longer than the age of the Universe
$\sim 10^{10} {\rm year}$.
This condition requires $\ell < 10^{-4}{\rm mm}$ (and hence $N< 10^{27}$).


{\it Effective pressure equations:}

The other non-trivial components of Israel junction condition dominate the acceleration/deceleration 
of our brane Universe.
As well as the effective energy density Eq. (\ref{eff_e}),
by integrating over the KK directions,
the effective four-dimensional pressure along the ordinary three-space
$P$ and that along the KK directions $\tilde Q$ are introduced
by integrating the ($(D-1)$-dimensional pressures $p$ and $q$
over the KK dimensions
\begin{eqnarray}
&& P:=p
\frac{Q^{(D-5)/(D-6)}
                  \Omega_{(D-5)}}
            {(1-a^4)^{(D-5)/(D-6)}},
\nonumber \\
&&\tilde Q
:= q \frac{Q^{(D-5)/(D-6)}
                  \Omega_{(D-5)}}
            {(1-a^4)^{(D-5)/(D-6)}},
\end{eqnarray}
respectively.
As for the effective energy density $\Sigma$,
those pressures are also decomposed into
the constant and time-dependent parts in general:
$P=P_0+P_1,\quad 
\tilde Q =\tilde Q_0 + \tilde Q_1$.
Note that the constant part is given by $P_0=\tilde Q_0=-\Sigma_0$.
By substituting the explicit expressions
for $F_{,a}/F$, ${\tilde B}_{,a}/{\tilde B}$ and $C_{,a}/C$
into  Eqs. (\ref{acc}) and (\ref{acc2}), we obtain 
\begin{widetext} 
\begin{eqnarray}
&& \frac{\ddot{a}}{a}
- \frac{(D-6)-a^4(5D-26)}{(D-6)(1-a^4)}
  \frac{\dot{a}^2}{a^2}
+\Big(\frac{D-6}{4}\Big)^2
 \frac{4\mu (1-a^4)^{2+2/(D-6)}}{a^4Q^{1+2/(D-6)}}
=-\frac{1}{2M_4^2}
  \frac{(1-a^4)^{3+2/(D-6)}}{1+\frac{D-2}{3(D-6)}a^4}
 \Big(1+\frac{\Sigma_1}{\Sigma_0}\Big)
 \Big(\Sigma_1+P_1\Big)
\nonumber\\
&&
\frac{\ddot{a}}{a}
+\frac{\dot{a}^2}{a^2}
 \frac{4a^4}{1-a^4}
+\Big[
1-\frac{4a^4}{(1-a^4)(D-6)}
+\frac{\frac{4\mu}{Q}}
   {a^4(1+\frac{2\mu}{Q}) -\frac{2\mu}{Q}}
\Big]
\Big(\frac{D-6}{4}\Big)^2
\frac{(1-a^4)^{2+2/(D-6)}}{a^4Q^{2/(D-6)}}
\big(a^4(1+\frac{2\mu}{Q}) -\frac{2\mu}{Q}\big)
\nonumber\\
&=&-\frac{1}{2M_4^2}
  \frac{(1-a^4)^{3+2/(D-6)}}{1+\frac{D-2}{3(D-6)}a^4}
 \Big(1+\frac{\Sigma_1}{\Sigma_0}\Big)
 \Big(\Sigma_1+\tilde Q_1\Big),\label{acc3}
\end{eqnarray}
\end{widetext}
where $M_4$ is defined in Eq. (\ref{pl2}).
In the early times $a\ll 1$ and
low energy scales $\Sigma_1/\Sigma_0 \ll 1$,
the external pressure equation can be reduced to the form 
\begin{eqnarray}
&& \frac{\ddot{a}}{a}
- \frac{\dot{a}^2}{a^2}
\approx
-\frac{1}{2 M_4^2}\big(\Sigma_1 + P_1
+\Sigma_{\rm DR} + P_{\rm DR}\big),
\end{eqnarray}
where the dark energy density and
pressure are given by Eq. (\ref{dr2})
$P_{\rm DR}=\Sigma_{\rm DR}/3$.
This is nothing but the result in the standard cosmology
except for the dark radiation type contribution in
the non-extremal case.

Hereafter, we focus on the extremal case $\mu=0$
and discuss the behavior of the pressure in the KK direction $\tilde Q_1$.
In the early times $a\ll 1$, we find
\begin{eqnarray}
\frac{1}{Q^{1/2}}
\Big(1 +\frac{\delta}{\Sigma_{\rm RS}}\Big)^2
\approx\frac{1}{6 M_4^2}
 \big(
 3P_1 -2\Sigma_1 -3\tilde Q_1
 \big).\label{cons}
\end{eqnarray}
Note that in the above Eq. (\ref{cons}),
the left-hand-side is constant.
Thus, in the right-hand-side, the combination $3P_1-2\Sigma_1-2\tilde Q_1$
must be totally time-independent.
Also, in the above derivation, it is assumed that
$\Sigma_1/\Sigma_0\ll 1$. 
For the realistic brane matter $P_1/\Sigma_0\ll 1$
should also be satisfied.
In order for Eq. (\ref{cons}) to be compatible,
$\tilde Q_1$ should be as large as $|\tilde Q_1|\sim \Sigma_0$
in contrast to $\Sigma_1$ and $P_1$.

One might think that
the junction conditions for the $(D-5)-$form field might restrict the
motion of the brane.
But now, it is not the case.
Only the non-trivial component of the $(D-5)$-form field strength
is $F^{y^1\cdots y^{D-5}}$, where
all the coordinates $y^a$ represent the dimensions of $S^{(D-5)}$, and
the product of the form field with the normal vector
$n_A F^{A_1\cdots A_{D-5}}$ vanishes. 
Thus all the junction conditions of the form become trivial.
In other words, the $(D-5)$-form field cannot couple with the
brane matter, if it is magnetic.


{\it Summary:}

We discussed cosmology in the brane Universe in
$D(>6)$-dimensional bulk spacetime in the context of
the {\it Kaluza-Klein (KK) braneworld} scheme,
i.e., to consider KK compactifications on the brane.

We consider the $D$-dimensional black three-brane solutions.
In the near horizon region the spacetime structure is approximately given by
${\rm AdS}_{5}\times S^{(D-5)}$ and
in the asymptotic infinity it approaches
the $D$-dimensional Minkowski.
We consider a brane motion from the near-horizon region toward the asymptotic infinity,
which induces the cosmology on the brane. 
We derive all the components of Israel junction condition,
by assuming exact cosmological symmetry on the (KK) brane.
The junction conditions for the $(D-5)$ form field do not restrict the brane motion.
In other words, the brane matter does not couple to the $(D-5)$-form field in the bulk.
After integrating over the compact KK dimensions,
we have derived the effective cosmological equations on the brane.
We find that irrespectively of components of the matter on the brane,
the brane cannot reach the asymptotic infinity of the bulk.
Thus, observers on the (KK) brane never see these $(D-5)$ extra dimensions.

In the early times, when the KK brane is moving in the near-horizon region,
the brane cosmology exactly coincides with that in the five-dimensional
Randall-Sundrum (RS II) model.
If the original black three-brane solution is near-extremal $\mu/Q\ll 1$,
the dark radiation type contribution arises.
It is natural since
the near horizon geometry is the product of five-dimensional 
AdS-Schwarzschild with $S^{(D-5)}$, which is similar to the RS model.
We also discussed experimental and observational bounds.

There would be various extensions of our considerations.
As one of the possible extensions, it would be important
to consider the cases without $Z_2$-symmetry across the brane,
which are rather generic in the higher dimensions.
In the present set-up, 
the physical size of our Universe approaches the maximal size.
This would be the disadvantage of our simplest model.
In order to avoid this problem and to obtain ever expanding 
brane Universe, it would be useful to consider the time-dependence
of the extra dimensions.

\section*{Acknowledgement}
The author wishes to thank the anonymous reviewer for his/her comments
and suggestions to improve the paper.
This work was supported in part by the Transregional Research Centre TRR33 
{\it The Dark Universe}.

\end{document}